\documentclass[referee]{aa}  
\usepackage[varg]{txfonts}
\usepackage{graphicx}
\usepackage{gensymb}
\usepackage{txfonts}
\usepackage[version=3]{mhchem}
\usepackage{hyperref}
\usepackage{amsmath}
\begin{document} 

\title{Destruction of interstellar methyl cyanide (\ce{CH3CN}) via collisions with \ce{He^{+.}} ions$^\dag$}


   \author{
        Luca Mancini \inst{1}
        \and
        Emília Valença Ferreira de Aragão  \inst{1}
        \and
        Fernando Pirani \inst{1}
        \and
        Marzio Rosi \inst{2}
        \and
        Noelia Faginas-Lago \inst{1}
        \and
        Vincent Richardson \inst{3, 4}
        \and
        Luca Matteo Martini \inst{4}
        \and
        Linda Podio \inst{5}
        \and
        Manuela Lippi \inst{5}
        \and
        Claudio Codella \inst{5}
        \and
        Daniela Ascenzi \inst{4}
          }

\institute{
Dipartimento di Chimica, Biologia e Biotecnologie, Università degli Studi di Perugia, via Elce di sotto, 8, 06123, Perugia, Italy
        \email{luca.mancini@unipg.it}\\
        \and
Dipartimento di Ingegneria Civile ed Ambientale, Università degli Studi di Perugia, via G. Duranti, Perugia, Italy
        \email{marzio.rosi@unipg.it}\\
        \and
Department of Physics, The Oliver Lodge, University of Liverpool, Oxford St, Liverpool, L69 7ZE, UK
        \email{Vincent.Richardson@liverpool.ac.uk}\\
        \and
Dipartimento di Fisica, Università di Trento, Via Sommarive 14, 38123 Trento, Italy
        \email{daniela.ascenzi@unitn.it}\\
        \and
INAF, Osservatorio Astrofisico di Arcetri, Largo E. Fermi 5, 50125 Firenze, Italy
        \email{linda.podio@inaf.it}
             }

   \date{Received 26th July, 2024; accepted XX XX, 2024}
 
  \abstract
    {\ce{CH3CN} (methyl cyanide) is one of the simplest and most abundant interstellar complex organic molecules (iCOMs), having been detected in young solar analogues, shocked regions, protoplanetary disks and comets. \ce{CH3CN} can therefore be considered a key species to explore the chemical connections between the planet forming disk phase and comets. For such comparison to be meaningful, however, kinetics data for the reactions leading to \ce{CH3CN} formation and destruction must be updated.}
    {Here we focus on the destruction of methyl cyanide through collisions with \ce{He^{+.}}. A combined experimental and theoretical methodology is employed to obtain cross sections (CSs) and branching ratios (BRs) as a function of collision energy, from which reaction rate coefficients $k(T)$ are calculated in the temperature range from 10 to 300 K.}
    {CSs and BRs are measured using a guided ion beam set-up. A theoretical treatment based on an analytical formulation of the potential energy surfaces (PESs) for the charge exchange process is developed. The method employs a Landau Zener model to obtain reaction probabilities at crossings between the entrance and exit PESs, and an adiabatic centrifugal sudden approximation to calculate CSs and $k(T)$, from sub-thermal to hyper-thermal regimes.}
   {$k(T)$ and experimental BRs differ from those predicted from widely-used capture models. In particular, the rate coefficient at 10 K is estimated to be almost one order of magnitude smaller than what reported in the KIDA database. As for BRs, the charge exchange is completely dissociative and the most abundant fragments are \ce{HCCN+/CCNH+}, \ce{HCNH+} and \ce{CH2+}.} 
  {Our results, combined with a revised chemical network for formation of \ce{CH3CN}, support the hypothesis that methyl cyanide in protoplanetary disks could be mostly the product of gas-phase processes rather than grain chemistry, as currently proposed. These findings are expected to have implications in the comparison of the abundance ratios of N-bearing molecules observed in disks with cometary abundance ratios.}


\titlerunning{\ce{He^{+.}} + \ce{CH3CN} collisions}
\authorrunning{L. Mancini et al.}
\maketitle


\section{Introduction}
Methyl cyanide (\ce{CH3CN}), also known as acetonitrile or cyanomethane, is a molecule of great astrochemical interest among interstellar complex organic molecules (iCOMs), as it has been routinely detected in several regions of the interstellar medium (ISM) at different stages of star formation. Its first detection dates back to 1971, when some rotational lines were reported toward Sagittarius B \citep{Solomon1971} and, since then, it has been detected towards dark clouds \citep{Matthews1983}, diffuse and translucent clouds \citep{Thiel2017, liszt2018chemical, Thiel2019}, massive star-forming regions \citep{purcell2006ch3cn}, and the Galactic centre \citep{zeng2018complex}. Focusing on young solar analogues, \ce{CH3CN} has been observed in Class 0 and I hot-corinos \citep{Cazaux2003, Taquet2015, Yang2021, Bianchi2022, Ceccarelli2023}, in shocked regions \citep{Codella2009}, and in planet-forming disks, where it has been firstly detected in 2015 \citep{Oberg2015} and then routinely observed in a series of surveys of protoplanetary disks \citep{Bergner2018, Kastner2018, Loomis2018, Loomis2020, Ilee2021}. 
\ce{CH3CN} has also been detected in comets \citep{Mumma2011}, including towards comet 67P/Churyumov-Gerasimenko in the context of the ESA-Rosetta mission \citep{LeRoy2015, Biver2019, Altwegg2019, Haenni2021}. 

Nitriles have a strong prebiotic relevance, since they are key intermediates in the formation of biomolecules such as amino acids and RNA precursors via reaction with water in a multi-step synthesis \citep{Sutherland2017}. For this reason, the presence of nitriles and water in comets, with \ce{CH3CN} ranging from $\sim 0.008$ to $0.054\%$ with respect to water \citep{Oberg2015, Biver2019}, is of particular interest and makes \ce{CH3CN} a key species to explore the chemical connections between protoplanetary disks and comets. However, for such comparison to be meaningful, astrochemical network databases such as KIDA \citep{KIDA2015} and UMIST \citep{UMIST2022} require accurate data on the reaction rates and BRs for the various \ce{CH3CN} formation and destruction pathways.

Historically, two main formation mechanisms for methyl cyanide have been proposed, one proceeding via grain-surface reactions and the other operative in the gas phase. The dominant grain-surface mechanisms are the ice-mediated \ce{CH3} + \ce{CN} association reaction and the hydrogenation of \ce{C2N} on the ice surface \citep{garrod2008complex}, while the radiative association reaction between \ce{CH3+} and \ce{HCN} represents the main formation route in the gas-phase \citep{Ceccarelli2023}. The network of gas-phase formation routes of \ce{CH3CN} has been recently extensively revised and updated \citep{Giani2023}, confirming the importance of the radiative association process and proposing new gas phase formation routes, such as \ce{CH3OH2+} + \ce{HNC}, \ce{CH3CNH+} + \ce{e-} and \ce{CH3CNH+}+ \ce{NH3}. 

In addition to interactions with photons and electrons \citep{Mezei2019}, the destruction mechanisms of \ce{CH3CN} are expected to be dominated by collisions with energetic ions, such as \ce{H+}, \ce{H3+}, \ce{HCO+} and \ce{He^{+.}}.
 While the reactions of \ce{H3+} and \ce{HCO+} lead mostly to non dissociative proton transfer (see the results of room temperature experiments in \cite{Mackay1976, Liddy1977} and the proposed rates in KIDA \citep{KIDA2015} and UMIST \citep{UMIST2022} databases), collisions with \ce{H+} and \ce{He^{+.}} are mostly destructive. This is because, for these species, the reactivity is dominated by highly exothermic charge transfer (CT) processes, thereby enabling dissociation.

In the case of \ce{He^{+.}} this is due to the large difference, 12.39 eV, between the recombination energy of \ce{He^{+.}} and the ionization energy (IE) of \ce{CH3CN}. Collisions with \ce{He^{+.}} have been proposed as important pathways for the decomposition of iCOMs ranging from \ce{CH3OCH3} and \ce{HCOOCH3} \citep{cernuto2017experimental,Cernuto2018,CeccarelliAscenzi2019} to \ce{CH3OH} \citep{richardson2022fragmentation}.
However, for \ce{CH3CN}, while the reaction with \ce{H+} has been experimentally studied at room temperature \citep{Smith1992,KIDA2015,UMIST2022}, to the best of our knowledge, no previous experimental or theoretical studies have been carried out for the reaction with \ce{He^{+.}}, with the rates and BRs reported in the aforementioned astrochemical databases referring to predictions from capture models and chemical intuition. 

From a more general perspective, barrierless elementary chemical processes (mostly driven by ions, radicals or atoms/molecules in long lived electronically excited states, \textit{e.g.} Penning processes) can occur under both sub-thermal conditions, such as those present in cold interstellar environments, and hyper-thermal conditions occurring in natural and laboratory plasmas, flames, combustion processes. The detailed characterization of such processes is not only relevant for fundamental studies, but also for applications well beyond the context of astrochemistry. 
Ion-molecule studies can employ a wide range of experimental methods (from guided ion beams and ion traps \citep{Caselli2023iontrap} to the Rydberg-Stark merged-beam approach \citep{Merkt2023Chimia, Merkt2024} and innovative techniques based on the use of Coulomb crystals \citep{Willitsch2023, Willitsch2023PCCP, Willitsch_NatChem2024, Willitsch2024, Lewandowski2024, Lewandowski2024_1}) to measure either partial/total reactive cross-sections as a function of collision energy, or rate constants as a function of temperature over a wide temperature range. 

In this paper we report partial and total \ce{He^{+.}} + \ce{CH3CN} reactive cross sections (CSs) measured as a function of collision energy over a range of around three orders of magnitude (from few tens up to more than $10^4$ meV). The analysis of experimental results, carried out by a synergistic theoretical framework, exploits an adiabatic centrifugal sudden approximation for the capture dynamics of reagents in the entrance channels \citep{Clary1990, maciel2006} and assumes that reactivity is triggered by non-adiabatic effects, occurring at the crossings between entrance and exit channels, described within a Landau Zener treatment. The analytical formulation of the Potential Energy Surfaces (PESs) for the entrance and exit channels, coupled with the treatment of the collision dynamics, allows the prediction of CSs and rate constants from sub-thermal up to hyper-thermal conditions.

Results from experiments probing the dynamics of reactive collisions, as in the present work, represent crucial tests for the simple/approximate methods often employed to describe reaction probabilities of ion-molecule processes occurring under sub-thermal conditions. In particular, capture theory predictions, determined exclusively by the strength of long range attraction, are likely to over-estimate the value of the rate coefficients in all cases where the short-range part of the interaction potential is important, such as in the presence of crossings between multiple nearby potential energy surfaces (PES) or when the dynamics is not adiabatic \citep{tsikritea2022capture}. Therefore, for realistic estimates of kinetic rate constants and product BRs under conditions of relevance from astrochemistry to laboratory plasmas, it is necessary to derive the underlying PESs and to properly describe the collision dynamics. Finally, as emphasized by the present investigation, comparison of  calculations with experimental results
is essential to test, and eventually refine, the accuracy of theoretical models.

Section \ref{sec:ExpMeth} summarizes basic details of the adopted experimental technique,  while Section \ref{sec:theo_meth} provides the framework of the theoretical methodologies. Section \ref{sec:res} details the obtained results, their analysis, discussion and relevance for astrochemical modelling. Conclusions follow in Section \ref{sec:conclusions}.

\section{Experimental Methodology}
\label{sec:ExpMeth}
The experimental data presented here have been collected using the GEMINI (Gas-phase Experiment for Measurements on Ion Neutral Interactions) setup at the University of Trento. GEMINI is a tandem mass spectrometer composed of two octopoles (O) and two quadrupole mass filters (Q) in a O-Q-O-Q configuration that allows for the investigation of bimolecular reactions of mass-selected ions (see \citep{Ascenzi07, Franceschi07} for details). Total, $\sigma_T$, and partial CSs as well as BRs have been recorded as a function of the collision energy in the centre of mass (CM) frame ($E$) by measuring the yields of both parent and product ions.

\ce{He^{.+}} ions are produced in the ion source chamber by electron ionization (electron energies in the 70-80 eV range) of \ce{He}, which is introduced at pressures of $\sim10^{-6}$ mbar. Following ionization, ions pass through the first octopole, which acts as a ion guide, before being mass selected by the first quadrupole. Reactions occur in the second octopole, which is surrounded by a scattering cell. Here, a $\sim$1\% mixture of \ce{CH3CN} in \ce{Ar} has been chosen as the neutral reactant in order both to stabilise the vapour pressure of \ce{CH3CN} and to keep any secondary reaction(s) at a reduced level. The mixture is introduced at variable pressures in the range of 1.0 x10$^{-8}$ to 1.3 x10$^{-4}$ mbar, which is monitored by a spinning rotor gauge (SRG2 MKS Instruments, MA. USA). 

The collision energy in the laboratory frame is dependent on both the reagent ion charge (+1 in the case of this work) and the difference between the ion source and reaction cell potentials. The retarding potential method \citep{teloy} was employed to determine the maximum of the first derivative of the reagent ion yield, which defines the zero of the kinetic energy. In this way, we have estimated an average reagent ion beam FWHM of $\sim 1$ eV in the laboratory frame, equivalent to $\sim 0.9$ eV in the center-of-mass (CM) frame. By varying the potentials of the second octopole and all subsequent optics, we are therefore able to scan a collision energy in the CM frame ($E$) in the range from $\sim0.03$ to $\sim10$ eV .

\section{Theoretical Methodology}
\label{sec:theo_meth}
\begin{figure}[h]
\centering
  \includegraphics[width=13cm]{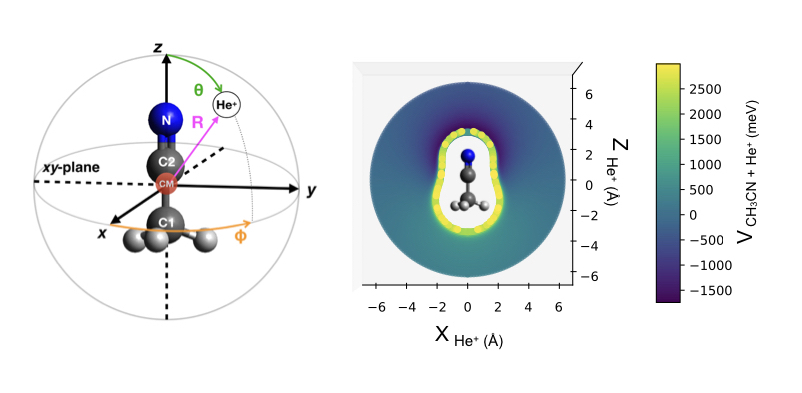}
  \caption{\textit{Left:} coordinate system used for the description of the \ce{He^{+.}} + \ce{CH3CN} interaction. Due to the cylindrical symmetry of \ce{CH3CN} and \ce{CH3CN^{+*}} (in our formulation, where the \ce{CH3} group is treated as a single effective atom), the PES depends solely on the polar coordinates $R$ and $\theta$. \textit{Right:} 3D representation of the PES, in the $xz$ plane, in the entrance channel. The attractive and repulsive contributions are indicated in blue and yellow, respectively.}
  \label{fig:strutt}
 \end{figure}

The potential energy surfaces (PES) governing the dynamics of the reagents (entrance channel) and of products (exit channel) have been constructed using a well established phenomenological approach, previously applied to a large variety of systems \citep{cernuto2017experimental, Cernuto2018, CeccarelliAscenzi2019, marchione2022unsaturated, richardson2022fragmentation, falcinelli2023role, Giani2023}.
The PESs, given in analytical forms, are a function of $R$ (defined as the distance between \ce{He^{+.}} and the centre of mass of \ce{CH3CN} in the entrance channel, and as the distance between \ce{He} and the centre of mass of \ce{CH3CN+} in the exit channel) and the relative orientation of the reaction partners. In the present case, due to symmetry considerations, the relative orientation depends only on $\theta$, the angle formed by the $\boldsymbol{R}$ vector and the $C_{3v}$ symmetry axes of \ce{CH3CN} (taken as the $z$ axis). It should be noted that \ce{CH3CN} is a symmetric top rotor exhibiting rotational components both along and perpendicular to its $C_{3v}$ symmetry axis, where its high permanent electric dipole of 3.919 D \citep{NISTCCC, mudip_CH3CN} is aligned. Figure \ref{fig:strutt} (left panel) depicts the adopted coordinate system for the formulation of the PESs. 

Our approach identifies and describes the primary interaction components by exploiting empirical or semi-empirical formulas defined by a limited number of parameters, related to fundamental properties of the interacting partners. These include the electronic polarizability (that can be partitioned in atomic and group components), ionization potential, electron affinity, permanent charges, anisotropic charge distribution, and the shape and energy of the atomic and molecular orbitals involved in the electron transfer. 

The interested reader is referred to \citep{ManciniICSSA2024} for details of the PESs, including values of the above mentioned parameters. Here, we only highlight the importance of an analytical formulation of the PESs, allowing for the identification of a series of crossings between entrance and exit channels, where the reactive process is promoted by non-adiabatic couplings triggered by charge transfer effects. Such crossings typically occur at intermolecular distances shorter than those where the centrifugal barrier is at a maximum. It should be noted that, while they form the basis of our theoretical treatment, the presence of crossings are not considered in classical capture models.
Figure \ref{fig:strutt} (right panel) presents the 3D representation of the entrance channel PES. This visualization highlights the significant interaction anisotropy, which plays a crucial role in the stereo-dynamics of reagents within the entrance channels, as discussed in Sec. \ref{subsec:theo_dyn}. 

\subsection{Reaction dynamics treatment}
\label{subsec:theo_dyn}
The reaction dynamics are treated within the adiabatic centrifugal sudden approximation, where the centrifugal component, arising from the relative motion of the collision partners, is considered as an average (diagonal) term in the interaction potential formulation used in the calculation of CSs. Cuts of the strongly anisotropic PESs, performed at defined orientation angles of the interacting partners, provide effective adiabatic PE curves that are useful in visualizing, at each orientation, the crossings between entrance and exit channels. 

The transition probability at the crossing points between the entrance and exit PE curves is treated by adopting a similar strategy used in previous investigations \citep{cernuto2017experimental, Cernuto2018, CeccarelliAscenzi2019, richardson2022fragmentation}. In particular, only those crossings between the entrance and exit channels that are energetically accessible will be considered as contributing to the reaction probability. Such crossings are localized in the attractive region of the entrance channel and the non-adiabatic transition probability is evaluated within the Landau-Zener-Stückelberg approach \citep{landau1932theory, zener1932non, stuckelberg1932theory, nikitin1999nonadiabatic, nikitin2012theory}. Exploiting the simplicity of such treatment, we have been able to carry out CSs calculations from sub-thermal (few meV) to hyper-thermal  (10$^4$ meV) collision energies ($E$). 
Specifically, the probability $p_i$ for diabatic passage through the $i$-th crossing between cuts of the PESs, represented by two PE curves (at a fixed $\theta$ orientation angle), is given by:

\begin{equation}
    p_{i}(E,\theta, l) = \exp \bigg(\frac{-2\pi \cdot H_i(R_i,\theta)^2}{\hbar \cdot  v_R (l,E) \cdot \Delta_i}\bigg) 
\end{equation}

where $E$ and $l$ are the collision energy and the orbital angular momentum quantum number of the complex formed by the reagents, respectively. $H_i(R_i,\theta)$ represents the non-adiabatic coupling at the $i$-th crossing between the two PE curves, obtained at a specific angle $\theta$. $\Delta_i$  is the difference in slope between the entrance and exit curves evaluated at the value of the $R$ coordinate at the crossing $R=R_i$. The radial velocity, $v_R$, describing the approach of the reagents, is defined by: 

\begin{equation}
   v_{R}^2 = \frac{2}{\mu}\bigg[E\bigg(1-\frac{l(l+1)}{k^2 R_i^2}\bigg) -E_i \bigg]
\end{equation}

where $E_i$ represents the interaction energy at the crossing point ($R=R_i$) that needs to be evaluated with respect to the asymptotic energy state of the reagents. $\mu$ is the reduced mass of the \ce{He^{.+}}-\ce{CH3CN} system and $k$ is the wavenumber defined as $k=$
\( \displaystyle \frac{\mu \cdot v_{R}}{\hbar}\). Following the guidelines given in literature \citep{olson1971estimation, gislason1975multiple, pirani2000coupling, candori2001structure}, $H_i^2(R_i,\theta)$ has been formulated as:

\begin{equation}
   H_i(R_i,\theta) = A\cdot R_i \cdot \bigg(1+ P_2(\cos\theta)\bigg) \cdot e^{-\alpha R_i}
\end{equation}

where $P_{2}(\cos\theta)$ is the second order Legendre polynomial. The parameter $\alpha$, depending on the IE and electron affinity (EA) of reagents comprising an electron donor (\ce{CH3CN}) and an electron acceptor (\ce{He^{+.}}), has been assumed equal to 2.50 \AA$^{-1}$ \citep{pirani2000coupling}. 
Therefore, only the pre-exponential $A$ value has been adjusted in order to reproduce the energy dependence of the total integral reactive CS, $\sigma_{T}(E)$, that has been measured experimentally (see Sec. \ref{sec:res}). The obtained $A$ value is $2.0\times10^5$ meV. 

At each orientation angle $\theta$, the contribution to the total integral CS for the charge transfer evaluated at a given $E$ value is considered as a sum of contributions from each $l$ value:

\begin{equation}
   \sigma(E,\theta) = \frac{\pi}{k^2}\sum_{l=0}^{l_{max}}(2l+1) \cdot P_i(E,\theta, l)
   \label{eq:sigma}
\end{equation}
\noindent 
where the formation probability of \ce{CH_3CN^{.+*}}, $P_{i}(E, \theta, l)$, is expressed in terms of the previously defined $p_i(E,\theta, l)$:

\begin{equation}
   P_{i}(E, \theta, l) = (1-p_{i}(E,\theta, l)) \cdot (1+p_{i}(E,\theta, l))
   \label{eq:bigP}
\end{equation}

Moreover, $l_{max}$ represents the maximum value of $l$ for which the system is able to reach the crossing point by overcoming the centrifugal barrier in the entrance channel. For high collision energies this is always the case and $l_{max}$ is given by:

\begin{equation}
   l_{max} = k R_i \sqrt{1-\frac{E_i}{E}}
\end{equation}
\noindent
that represents the maximum value of $l$ for which $v_R$ is real at the the $i$-th crossing point.

\section{Experimental results and theoretical implementation}
\label{sec:res}
\subsection{Experimental results: BRs and comparisons with other ionization methods for \ce{CH3CN}}

The main ionic products of the reaction of \ce{He^{.+}} with \ce{CH3CN} are those at \textit{m/z} 14, 28 and 39, while less intense products are observed at \textit{m/z} 13, 15, 27, 38 and 40. 
Notably, as we do not observe a product at \textit{m/z} 41, corresponding to the formation of \ce{CH3CN^{.+}}, we conclude that the electron transfer process is completely dissociative.  
However, we do note additional peaks at \textit{m/z} 42 and, though less intense, at \textit{m/z} 54, which are identified as products of the secondary reaction of the most prominent primary ions with excess \ce{CH3CN}.

The extent of secondary reaction has been minimized by using a mixture of \ce{CH3CN} diluted with \ce{He}, but it was impossible to completely eliminate secondary reactions due to the high reactivity of \ce{CH3CN} (having a large proton affinity of 8.076 eV \citep{NISTchem}) with many hydrocarbon and N-containing hydrocarbon ions via proton transfer. The presence of secondary reactions in ion reactivity studies with \ce{CH3CN} has been noted previously using quite different experimental set-ups \citep{Lewandowski2021, Lewandowski2024}. 

The assignments for the observed major primary products and their corresponding reaction energetics, \textit{i.e.} $\Delta$H$^{\circ}$(298 K) given in eV as calculated from available literature data \citep{ATcT,NISTchem}, are as follows: 
\begin{align}
\ce{
&He^{.+} + CH3CN & \rightarrow \hspace{2pt} & CH2^{.+} (\textit{m/z}\hspace{2pt} 14) + HCN/HNC + He  &-9.57/-8.91 \label{eq:2}\\ 
&               & \rightarrow \hspace{2pt} & HCNH^{+} (\textit{m/z}\hspace{2pt} 28) + CH^{.} + He   & -9.34 \label{eq:3}\\ 
&               & \rightarrow \hspace{2pt} & HCCN^{.+}/CCNH^{.+} (\textit{m/z}\hspace{2pt} 39) + H2 + He &-9.63/-8.92 \label{eq:4}
	}
\end{align}
Similarly, the assignments for the observed minor primary products are as follows: 
\begin{align}
\ce{
&He^{.+} + CH3CN & \rightarrow \hspace{2pt} & CH^{+} (\textit{m/z}\hspace{2pt} 13) + H2CN^{.}/HCNH^{.} + He  &-6.06/-5.71 \label{eq:5}\\ 
&               & \rightarrow \hspace{2pt} & CH3^{+} (\textit{m/z}\hspace{2pt} 15) + CN^{.} + He   & -9.44 \label{eq:6}\\ 
&               & \rightarrow \hspace{2pt} & CN^{+} (\textit{m/z}\hspace{2pt} 26) + CH3^{.} + He & -5.32 \label{eq:7}\\
&               & \rightarrow \hspace{2pt} & HCN^{.+}/HNC^{.+} (\textit{m/z}\hspace{2pt} 27) + CH2 + He & -6.35/-7.29 \label{eq:8}\\
&               & \rightarrow \hspace{2pt} & CCN^{+} (\textit{m/z}\hspace{2pt} 38) + H2 + H^{.} + He & -5.21 \label{eq:9}\\
&               & \rightarrow \hspace{2pt} & CH2CN^{+} (\textit{m/z}\hspace{2pt} 40) + H^{.} + He & -10.10 \label{eq:10}
}
\end{align}
Finally, the secondary reactions giving products at \textit{m/z} 42 and \textit{m/z} 54 are identified as:
\begin{align}
\ce{
& HCCN^{.+}/CCNH^{.+} + CH3CN & \rightarrow \hspace{2pt} & CH3CNH^{+} (\textit{m/z}\hspace{2pt} 42) + CCN  &-0.89/-1.60 \label{eq:11}\\ 
& HCNH^{+} + CH3CN    & \rightarrow \hspace{2pt} & CH3CNH^{+} (\textit{m/z}\hspace{2pt} 42) + HCN   & -0.73 \label{eq:12}\\ 
& CH2^{.+} + CH3CN    & \rightarrow \hspace{2pt} & CH3CNH^{+} (\textit{m/z}\hspace{2pt} 42) + CH & -0.51 \label{eq:13}\\
& HCN^{.+}/HNC^{.+} + CH3CN  & \rightarrow \hspace{2pt} & CH3CNH^{+} (\textit{m/z}\hspace{2pt} 42) + CN & -2.62/-1.68 \label{eq:14}\\
& CH2CN^{+} + CH3CN  & \rightarrow \hspace{2pt} & H2CCNCH2^{+} (\textit{m/z}\hspace{2pt} 54) + HCN & -1.93 \label{eq:15}
	}
\end{align}

Since the secondary product at \textit{m/z} 42 (\ce{CH3CNH+}) can originate from more than one primary reactant ion, the different reaction rates for the respective secondary processes must be considered in order to obtain accurate BRs for the various primary reaction channels. 

For reaction (\ref{eq:11}), reaction rates in literature are in the 3.00-3.54x10$^{-9}$ cm$^{3}\cdot$s$^{-1}$ range \citep{Franklin1966, Vogt1975, Anicich03}, with the main channel being proton transfer \citep{Gray1968, Wincel1988}. 
For reactions (\ref{eq:12}) and (\ref{eq:13}), the proton transfer process is the only observed channel, with reported rate constants of 3.80x10$^{-9}$ cm$^{3}\cdot$s$^{-1}$ \citep{McEwan1989} and 1.76x10$^{-9}$ cm$^{3}\cdot$s$^{-1}$ \citep{Franklin1966, Anicich03}, respectively. While no literature reaction rate information has been found for reaction (\ref{eq:14}), a direct comparison of the estimated rates for reactions (\ref{eq:14}) and (\ref{eq:12}) obtained using the Su-Chesnavich capture rate model \citep{Su1982, tsikritea2022capture} indicate that the rates for the two reactions are very similar. For this reason, the literature reaction rate for reaction (\ref{eq:12}) of $k=$3.80x10$^{-9}$ cm$^{3}\cdot$s$^{-1}$ is assumed for reaction (\ref{eq:14}). 

For the secondary product observed at \textit{m/z} 54 (\ce{CH3CNH+}), the only pathway requiring consideration is the one described by reaction (\ref{eq:15}) \citep{Oldham1999, Dechamps2007, Ascenzi2012}, for which reported rate constants are in the range 1.78-2.09x10$^{-9}$ cm$^{3}\cdot$s$^{-1}$ \citep{Franklin1966, Vogt1975}. 

\begin{figure}[h]
\centering
  \includegraphics[width=12cm]{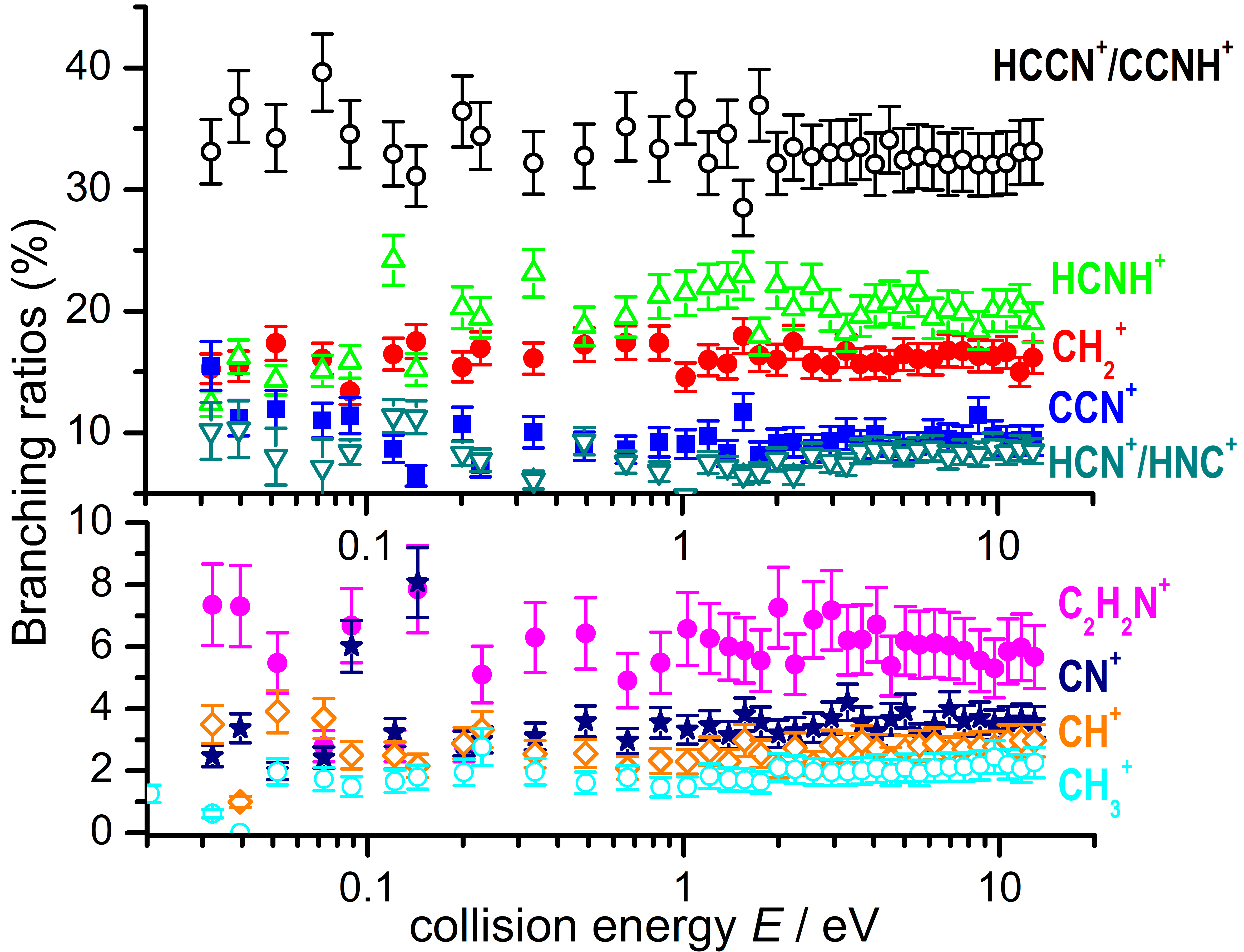}
  \caption{Experimental BRs for the reaction of \ce{He^{+.}} with \ce{CH3CN} as a function of collision energy.}
  \label{fig:expBRs}
 \end{figure}

\begin{figure}[h]
\centering
  \includegraphics[width=12cm]{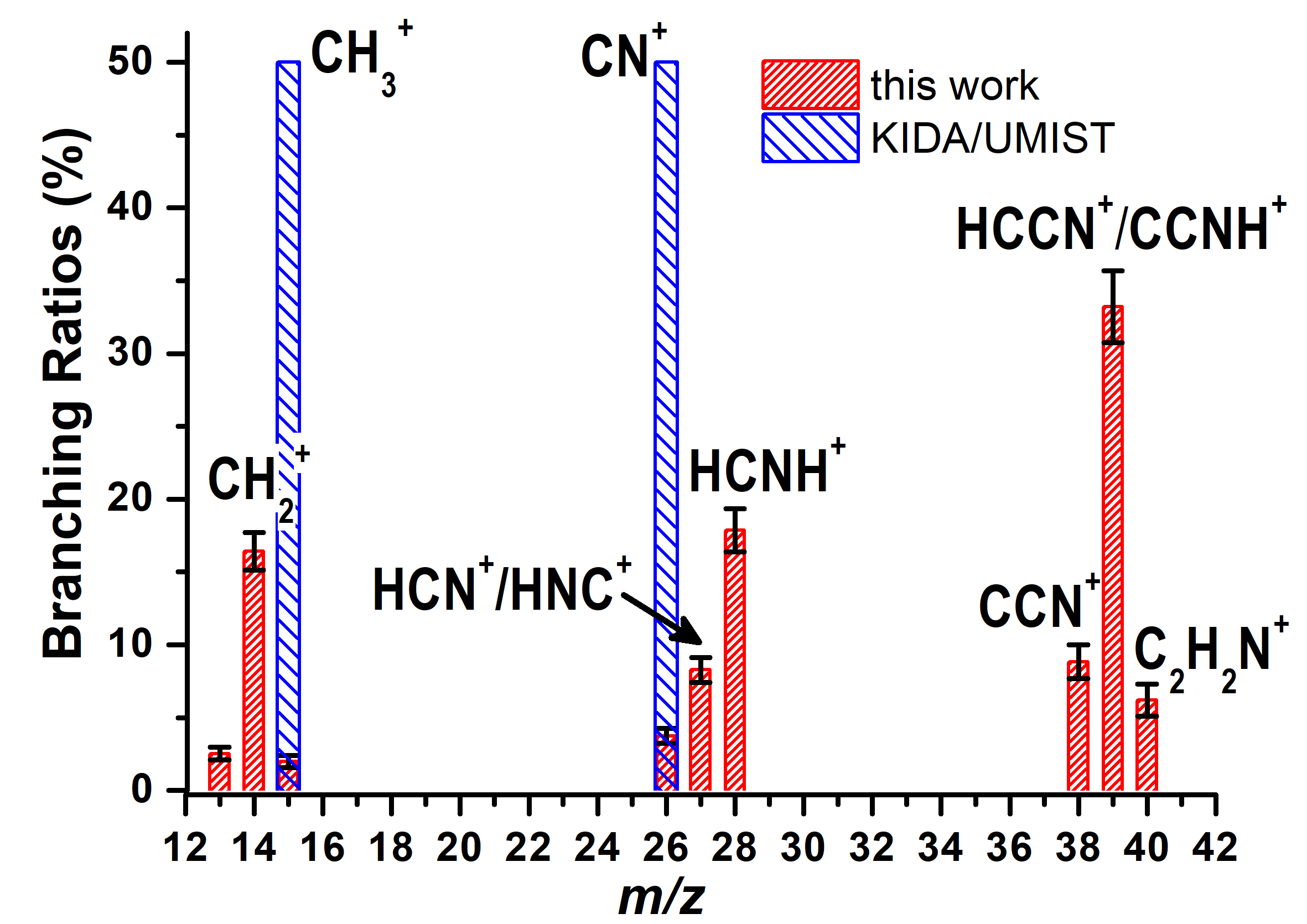}
  \caption{Branching ratios (BRs) for the reaction between \ce{He^{+.}} and \ce{CH3CN} as obtained in present experiments (red bars) and in literature databases \citep{UMIST2013, KIDA2015} (blue bars).}
  \label{fig:BRs}
 \end{figure}

The corrected primary channel BRs for the reaction of \ce{He^{+.}} with \ce{CH3CN}, measured as a function of the collision energy ($E$), are shown in Figure \ref{fig:expBRs}, while a comparison with BRs currently available in the astrochemical databases \citep{UMIST2013, KIDA2015} is shown in Fig. \ref{fig:BRs}. We note that the measured BRs are approximately constant over the studied collision energy range, spanning multiple orders of magnitude. 

It is instructive to compare the BRs for \ce{CH3CN} ionization measured by collision with \ce{He^{.+}} ions with other astrochemically relevant ionizing agents, namely electrons and photons.
The dissociative electron ionization of \ce{CH3CN} has been studied previously \citep{Harland1985, Price2019}, with relative precursor-specific partial ionization cross-sections for various fragment ions following dissociative single electron ionization at electron energies from 30 to 200 eV having recently been reported \citep{Price2019}. The measured BRs were approximately independent of the electron energy used, with the primary product being that at \textit{m/z} 40 (\ce{CH2CN+}) with a BR of 0.42 at an electron energy of 40 eV. The next most significant channels were those at \textit{m/z} 39 (\ce{HCCN+}/\ce{CCNH+}), 38 (\ce{C2N+}) and 14 (\ce{CH2+}), with respective BRs of 0.17, $\sim0.1$ and $\sim0.1$ at the same electron energy of 40 eV. 

Overall, the combined BR for the \textit{m/z} 38, 39 and 40 products, \textit{i.e} those where the C-C-N spine remains intact, and equal to $\sim$0.7, is higher than the $\sim$0.5 value obtained in the current work. This is indicative of a "softer" ionization, where \ce{H}- and \ce{H2}-ejection processes dominate. By contrast, while such processes are still significant in the case of the \ce{He^{.+}} plus \ce{CH3CN} reaction, we observe a higher contribution from processes involving C-C or C-N bond cleavage. This difference is expected to arise from the differing nature of the ionization processes involved, which in the present case lead to the generation of a highly electronically excited state of the \ce{CH3CN} radical cation, with further consideration given in Section \ref{sec:crossings}.

The ionization, isomerization and dissociation of methyl cyanide upon photon absorption in the gas phase has been extensively investigated \citep{Turner1970, Asbrink1980, Holland1984, Gochel1992, Holland2006, Huang2007, Schwell2008, Kukk2009, Polasek2016, Boran2017, McDonnell2020}. Relative and/or absolute photoionization CSs for the formation of both molecular and fragment ions have been measured in the VUV region, from the ionization threshold at 12.20 eV up to higher energies where inner shell photoionization events are possible. Additionally, the electron stimulated non-thermal ion desorption from an acetonitrile ice upon bombardment with high energy electrons (in the range 1000-2300 eV) has been reported \citep{Ribeiro2015}.

The most relevant study to compare with the results of the present work is \citep{Kukk2009}. There, the dissociation of acetonitrile following resonant core excitations was investigated by recording fragment ion mass spectra in coincidence with the resonant Auger electrons, emitted in the decay process of the core-excited states, using the photoelectron–photoion coincidence (PEPICO) technique. This technique allows for the investigation of molecular fragmentation of specific ionic states generated by participator and spectator Auger decays. Interestingly, in the spectra recorded in coincidence with electrons having a binding energy above $\sim22-23$ eV (similar to the IE of He) the \ce{CH3CN+} parent cation is almost entirely absent from the ion mass spectrum. Correspondingly, the observed fragments are dominated by \ce{CH_{n}^{+}} (with $n=1-3$), \ce{CNH_{n}^{+}} (with $n=0-2$) and \ce{H_{n}CCN^{+}} (with n=0-2), and the fragmentation pattern bears a close resemblance to that we observe by collisions with \ce{He^{+.}}. Furthermore, \citep{Kukk2009} shows that the \ce{CH3+}/\ce{CH2+} ratio is highly sensitive to the initial molecular ionic state, with the low ratio observed in our work ($\sim0.12$) being consistent with electron binding energies in the 24-27 eV range.

A previous combined experimental and theoretical study explored the isomerization and dissociation pathways accessible following laser photoionization of methyl cyanide \citep{McDonnell2020}. At the B3LYP/6-31++G(d,p) level of theory, pathways corresponding to reactions (\ref{eq:2})-(\ref{eq:8}) were identified and, as also established elsewhere \citep{Huang2007, Polasek2016}, isomerization through H migration leads to the stable linear (\ce{CH2CNH^{.+}}) and cyclic (\ce{c-CHCHNH^{.+}}) isomers, both of which have distinctive fragmentation pathways. For instance, dissociation into \ce{CH2^{.+}} plus \ce{HNC} is energetically favoured from the \ce{CH2CNH^{.+}} isomer, while \ce{H2}-ejection from \ce{CH3CN^{.+}} to give \ce{HCCN^{.+}} is energetically preferred to the equivalent ejection from \ce{CH2CNH^{.+}} to yield \ce{CCNH^{.+}}.

\subsection{Experimental results: CSs as a function of collision energy and comparisons with capture models}
\label{sec:expres1}
\begin{figure}[h]
\centering
  \includegraphics[width=12cm]{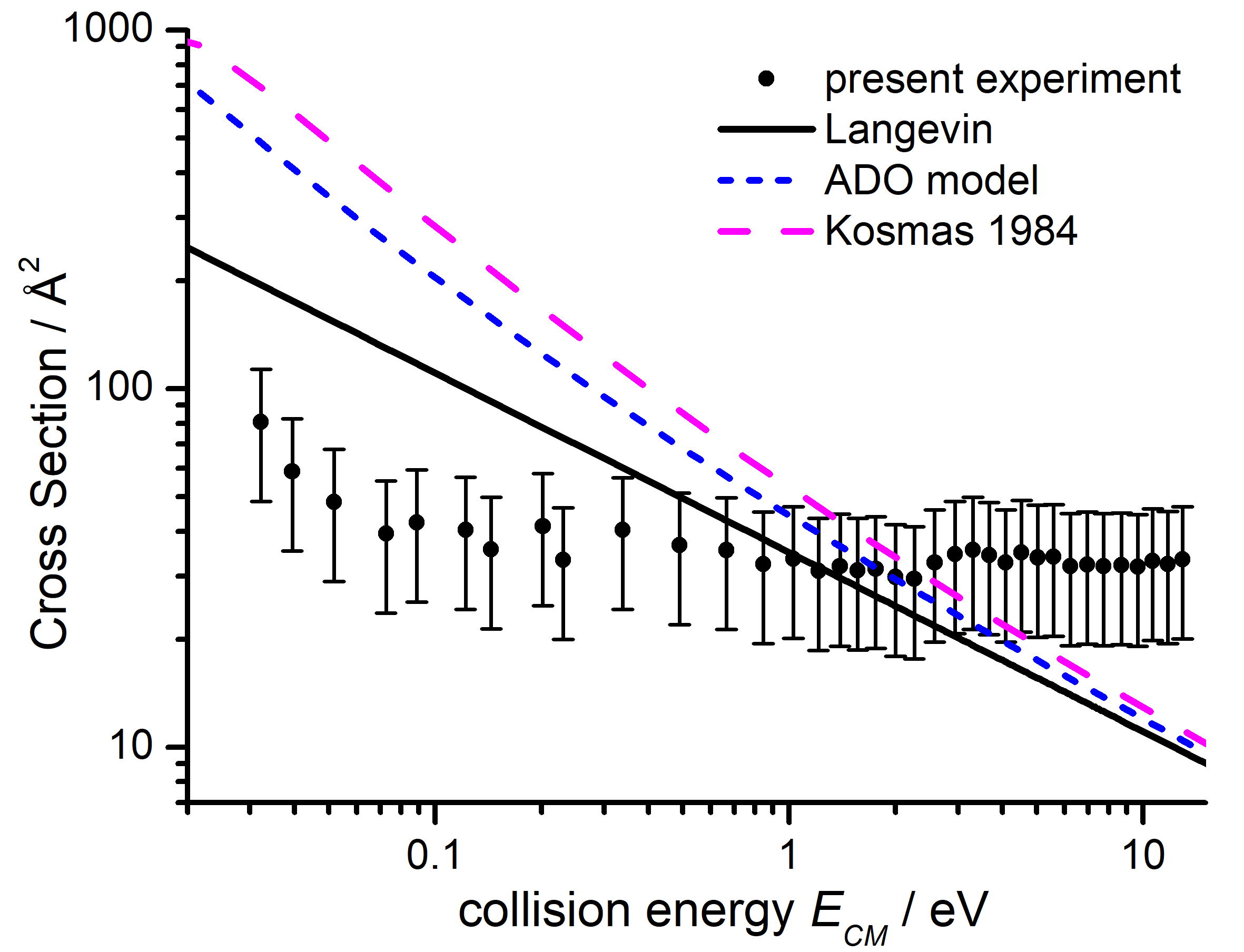}
  \caption{Total CSs for the electron exchange reaction of \ce{He^{+.}} with \ce{CH3CN} as a function of collision energy ($E$). The \textit{black filled circles} are the total relative CSs, rescaled as detailed in the text. Lines represent total CSs as estimated from the following capture models: Langevin (\textit{black solid line}), averaged dipole orientation (ADO) (\textit{blue short dashed line}) \citep{Su1975, Su1978, tsikritea2022capture} and the average $cos\theta$ approach \citep{Kosmas1984}(\textit{magenta dashed line}). Error bars on the total relative CSs are $\pm40\%$. 
  }
  \label{fig:TotCS}
 \end{figure}

Prior to performing the reaction dynamic treatment, we present a comparison between the experimental total CS trend as a function of the collision energy and those predicted by traditional capture models. Total CSs have been compared with those from capture models \citep{tsikritea2022capture} using the region with intermediate $E$ values (between 1-2 eV) for normalization, with results shown in Figure \ref{fig:TotCS}.

Notably, the Langevin model describing the interaction between a point-charge ion and a polarizable but non polar neutral (black solid line in Figure \ref{fig:TotCS}) is not expected to reproduce the probability of capture between an ion and a polar molecule. In order to provide a more realistic model, the average dipole orientation (ADO) theory, developed by Bowers and coworkers \citep{Su1975, Su1978}, additionally considers the ion-permanent dipole interaction. To account for the average orientation of the dipole of the neutral species with respect to the ion, a scaling parameter ($c=\overline{cos\theta}$) is introduced. The parameter, with values ranging from 0 to 1, depends on the dipole moment and on the polarizability of the neutral. Using the parametrization reported in \citep{Su1975} a $c$ value of 0.253 is used for the present system (blue short dashed line). 
A similar treatment, the so called average $cos\theta$ method proposed by \citep{Kosmas1984}, 
follows the line of the ADO theory, but gives expressions for the CSs as a function of both the relative kinetic energy of the colliding partners and the rotational energy of the polar molecule. The magenta dashed line is calculated using this model adapted to our experimental configuration, where the \ce{CH3CN} is in a collision cell at a fixed $T$. We therefore assume an average rotation energy equal to 0.039 eV, \textit{i.e.} $3/2kT$ with T=305 K (due to RF heating inside the octupole ion guide). 

In all cases, the calculated CSs exhibit a sharp decrease with increasing $E$, in contrast to the experimentally observed trend. Such inconsistencies between predictions and experimental findings highlight the inadequacy of capture models for treating the charge exchange process between \ce{He^{.+}} and \ce{CH3CN} in the explored collision energy range.

\subsection{Crossing points between entrance and exit channels}
\label{sec:crossings}
The collision energy (spanning a range from $\sim0.03$ to $\sim10$ eV in the present experiment) promotes non-adiabatic transitions at the crossings between the entrance and exit channels that are located in a wide range of separation distances. 
Moreover, the large difference in the IE of the two species (IE(He)= 24.588 eV = EA(\ce{He^{.+}}) and IE(\ce{CH3CN}) = 12.20 $\pm$ 0.01 eV \citep{NISTchem, Holland1984, Gochel1992, Holland2006}) does not permit crossings between entrance and exit channels in the case of an electron ejected by either the $2e(\pi_{CN})$ HOMO of \ce{CH3CN}, see Figure 6 in \cite{ManciniICSSA2024}, or other outer valence molecular orbitals with binding energies up to $\sim17$ eV. 

In order to have effective crossings between entrance and exit PE curves, the electron must be removed from a molecular orbital in the inner valence region, \textit{i.e.} with a binding energy much higher than 12.20 eV. Photoelectron spectra \citep{Asbrink1980, Holland2006}, and electronic structure calculations \citep{ManciniICSSA2024} suggest that the inner valence molecular orbital $5a_1$, responsible for the broad peak observed in the photoelectron spectrum in the region 23-27 eV (see \citep{Asbrink1980} and Fig. 1 in \citep{Holland2006}), has the most comparable energy to the IE of He, and is therefore the MO involved in the electron exchange process. Assuming the IE of $5a_1$ as the proper asymptotic level for the exit channels, a sequence of crossings is obtained where non-adiabatic charge transfer effects are essential to trigger the ion-molecule reaction, as detailed in \citep{ManciniICSSA2024}. 

\begin{figure}[h]
\centering
  \includegraphics[width=13cm]{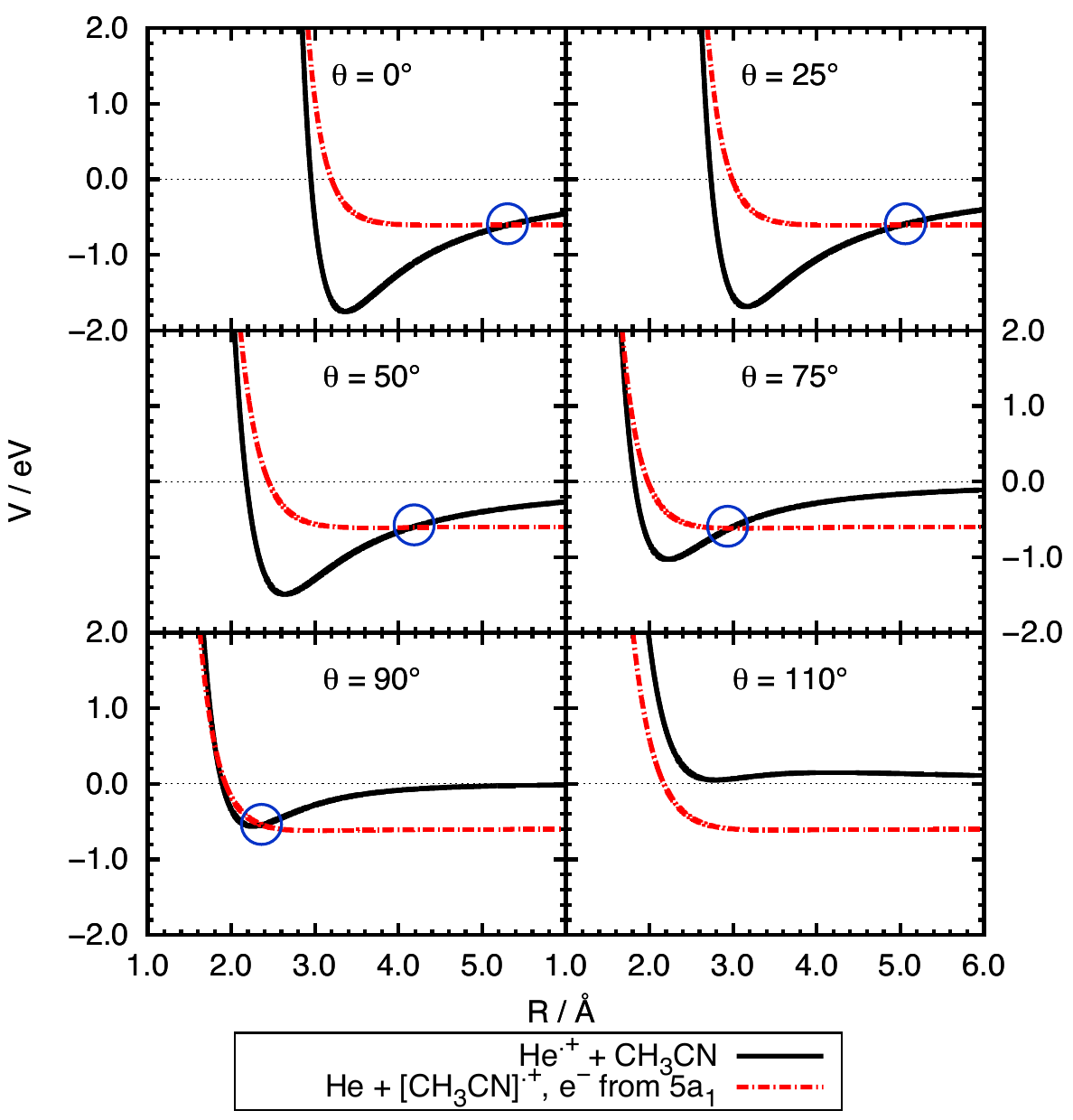}
  \caption{PES crossings: PE curves of the entrance \ce{He^{+.}} plus \ce{CH3CN} (solid black lines) and exit \ce{He} plus \ce{CH3CN^{.+*}} (dotted-dashed red lines) channels  for six different geometries determined by $\theta$ values equal to 0\degree, 25\degree, 50\degree, 75\degree, 90\degree\ and 110\degree\. The red dotted-dashed lines have been calculated assuming the removal of an electron from the inner valence orbital $5a_1$ of \ce{CH3CN}, thus forming the radical cation in an electronically excited state, but that such removal does not substantially change the geometry of the ion with respect to the neutral molecule. The position of crossings is highlighted with blue circles.}
  \label{fig:PEStagli2}
\end{figure}

In Figure \ref{fig:PEStagli2}, the curves obtained by PES cuts at six different geometries, corresponding to six different $\theta$ values from 0\degree\ to 110\degree, are reported. For each configuration, the solid black lines refer to the entrance channels (\ce{He^{+.}} plus \ce{CH3CN}), while the dotted-dashed red lines are the scaled exit potentials (\ce{He} plus \ce{CH3CN^{.+*}}) resulting from the removal of an electron from the $5a_1$ orbital of \ce{CH3CN}. The zero of the energy scale is taken to coincide with the value at $R\rightarrow \infty$ of the PE cut of the entrance channel. The red dotted-dashed lines have been calculated assuming that the removal of an electron from an inner valence orbital does not substantially change the geometry of the radical cation with respect to the neutral molecule. We also note that the most effective crossing points appear for $\theta$ varying from 0$\degree$ up to $\sim90\degree$, while no crossings occur for $\theta$ significantly larger than 90$\degree$.

It is relevant to note that, in the proposed PES, predicted interactions for three reference configurations (\textit{i.e.} $\theta=0$\degree, 90\degree and 180\degree) of the system in the exit channel, excluding ion-induced dipole induction contribution (defined as $V_{ind}$ in \cite{ManciniICSSA2024}), are in agreement within $\sim10$\% on the binding energies and 2-3\% on the equilibrium distances, with a full \textit{ab initio} treatment of the neutral complex He-\ce{CH3CN}, as reported in \citep{Dagdigian2022}.

\subsection{Reaction dynamics treatment: cross section calculations}

The analysis of the experimental data has been performed by adopting three different dynamical regimes of collision events. Since the PES driving the approach in the entrance channels is strongly anisotropic, exhibiting energy barriers that already exceed the average rotational energy of \ce{CH3CN} evaluated at room temperature at a separation distance of 20 \AA, the reagents undergo natural orientation effects induced by the strong electric field gradient associated with the anisotropic interaction. 

As a consequence, the collision complexes formed by the \ce{He^{+.}} ion and the \ce{CH3CN} polar molecule tend to re-orient themselves in angular cones ($cone_i$) confined within the attractive part of the PES, with the aperture of the acceptance cone changing with the collision energy $E$. 
In particular, considering the anisotropy of the potential well depths shown in Figure \ref{fig:PEStagli2}, it has been assumed that at low collision energies ($E \leq 0.1$ eV)  most of the molecules are confined in a restricted angular cone ($cone_0$) that exhibits a maximum opening angle $\theta=25$\degree. 
The opening angle increases up to $\theta = 45$\degree\ ($cone_1$)  for energies in the range $0.3 \leq  E \leq  0.6$ eV and again up to $\theta = 90$\degree\ for $E \geq 1.0$ eV ($cone_2$) .

The total charge exchange CS $\sigma_{T}(E)$ is then calculated as a sum of the contributions form the different angular cones, with relative weightings depending on the $E$ values. Accordingly:

\begin{equation}
   \sigma_{T}(E) = [\sigma_{cone_0}(E) \cdot f_{0}(E) + \sigma_{cone_1}(E) \cdot (1-f_{0}(E))] \cdot f_{1}(E) + \sigma_{cone_2}(E) \cdot (1-f_{1}(E))
\end{equation}

where $f_{i}(E)$ is a Fermi weight function: 
\begin{equation}
f_{i}(E) = \frac{1}{1+e^{\frac{E-E_i}{E_{ti}}}} 
\end{equation}

The values of the parameters $E_i$ and $E_{ti}$ define the collision energy at which the two combined  regimes of collision dynamics exhibit equal weightings and the rate at which the transition between the two calculation methods occurs, respectively. In the present study,  $E_0$ and $E_1$  have values of 0.25 and 0.65 eV, respectively, while $E_{t0}$ and $E_{t1}$ are 0.07 and 0.13 eV, respectively. 

At low $E$, reactivity is predominantly the result of collision events governed by the strongest long-range attractions, and primarily confined within $cone_0$. This leads to a significant attenuation of the reactivity, as all relevant crossing points between entrance and exit channels occur at large $R_i$. Furthermore, the non-adiabatic coupling $H_{i}(R_{i},\theta)$ is weak due to the limited overlap between the atomic and molecular orbitals involved in the electron exchange process. 

With increasing $E$, the reaction cone ($cone_1$) enlarges, and the crossing points shift to smaller $R_i$ values, where the strength of $H_{i}(R_{i},\theta)$ increases. At even higher $E$, the reaction cone further expands ($cone_2$), and the collision complexes can probe the entire semi-sphere of relative configurations where the PES is non-repulsive, \textit{i.e.} the region where the complete sequence of accessible crossing points occurs. Under such conditions, reactivity is influenced by the entire spectrum of crossing points. While crossings at $\theta$ angles close to 90\degree are associated with small $R_i$ and significant $H_{i}(R_{i},\theta)$, the contribution from these crossings is attenuated by the increased values of $E_i$ and, by extension, of the centrifugal potential, which limits the angular momentum $l$ values that allow approach to $R_i$. Moreover, the PES cuts shown in Figs. \ref{fig:PEStagli2} suggest that, for some $\theta$ values, additional crossings beyond those considered in the present analysis are likely open. However, their contribution is expected to be negligible since they occur in the repulsive region of the PES, and at short $R_i$ values, where the limitations due to the centrifugal contribution are greatest.

\begin{figure}[h]
\centering
  \includegraphics[width=12cm]{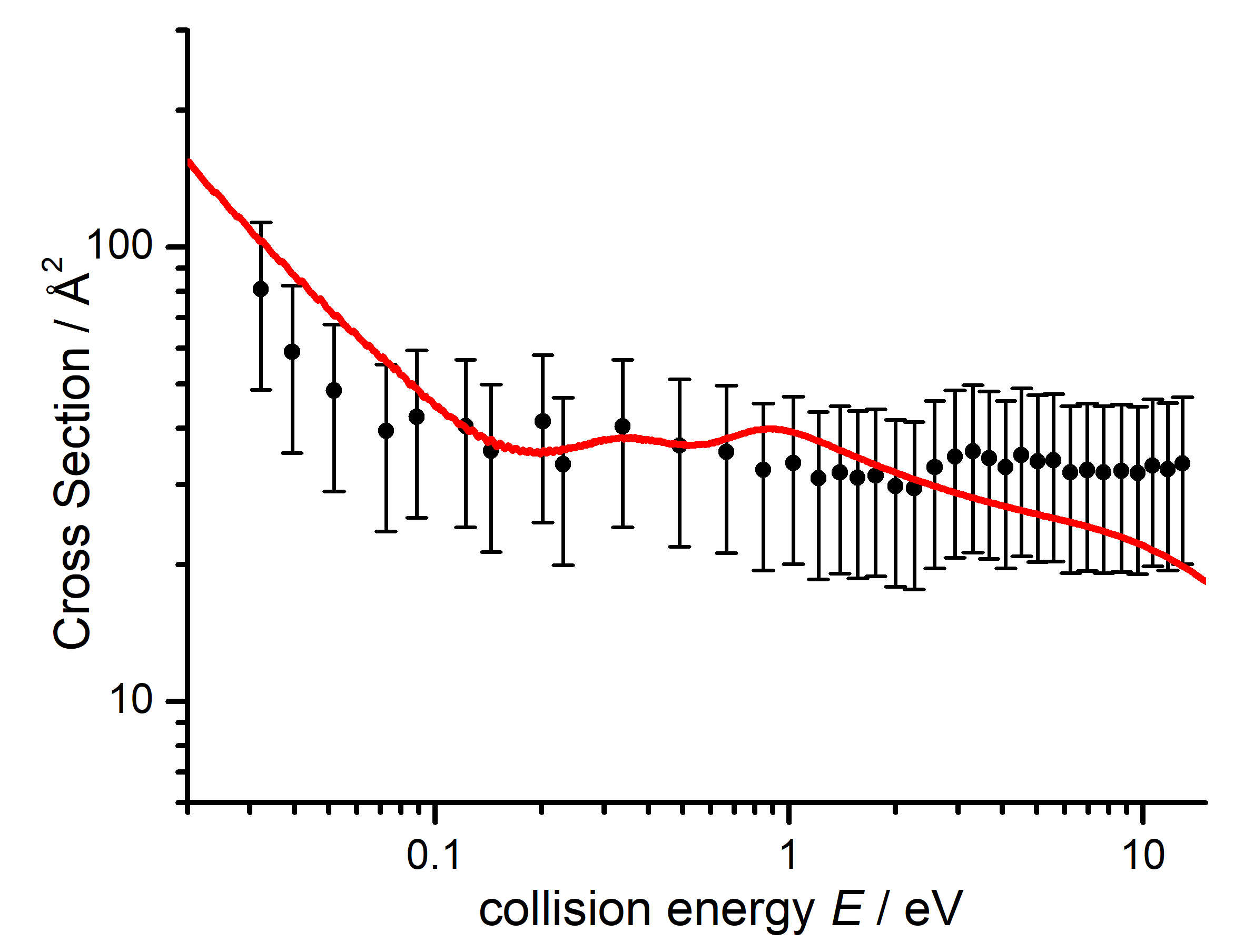}
  \caption{Comparison between the experimental total CSs for the electron exchange reaction of \ce{He^{+.}} with \ce{CH3CN} (black filled circles, already shown in Fig. \ref{fig:TotCS}) and the computed values according to the theoretical treatment described in the text and based on an improved Landau-Zener-Stückelberg approach (solid red line). Error bars on the total relative CSs are $\pm40\%$. 
  }
  \label{fig:TotCS_1}
 \end{figure}

Experimental and computed total CSs $\sigma_{T}(E)$ are compared in Figure \ref{fig:TotCS_1}. We note a good agreement in terms of the energy dependence of the measured CS in a wide range of $E$ values. According to our treatment, the barriers in the entrance PES tend to channel most of the collision complexes in the semi-sphere of the relative orientations where the attraction dominates and the size of the \ce{CH3} group, mostly responsible for the short range repulsion, plays a minor role. However, the effectiveness of the orientation tends to diminish at very high $E$ values. 

At low collision energies, the CSs evaluated using capture models (see Fig. \ref{fig:TotCS}) are higher than those calculated here, due to the fact that they do not consider the features of the crossing points and due to the selective role of the centrifugal barriers emerging in the entrance channels. At high collision energies, $i.e.$ when the probed intermolecular distances decrease, our treatment gives CSs that converge to the capture model ones, although the latter do not take into account the size of the \ce{CH3} group, since the interaction anisotropy is evaluated considering only the ion-permanent dipole interaction component.

\subsection{Rate constants and astrochemical relevance}

\begin{figure}[h]
\centering
  \includegraphics[width=12cm]{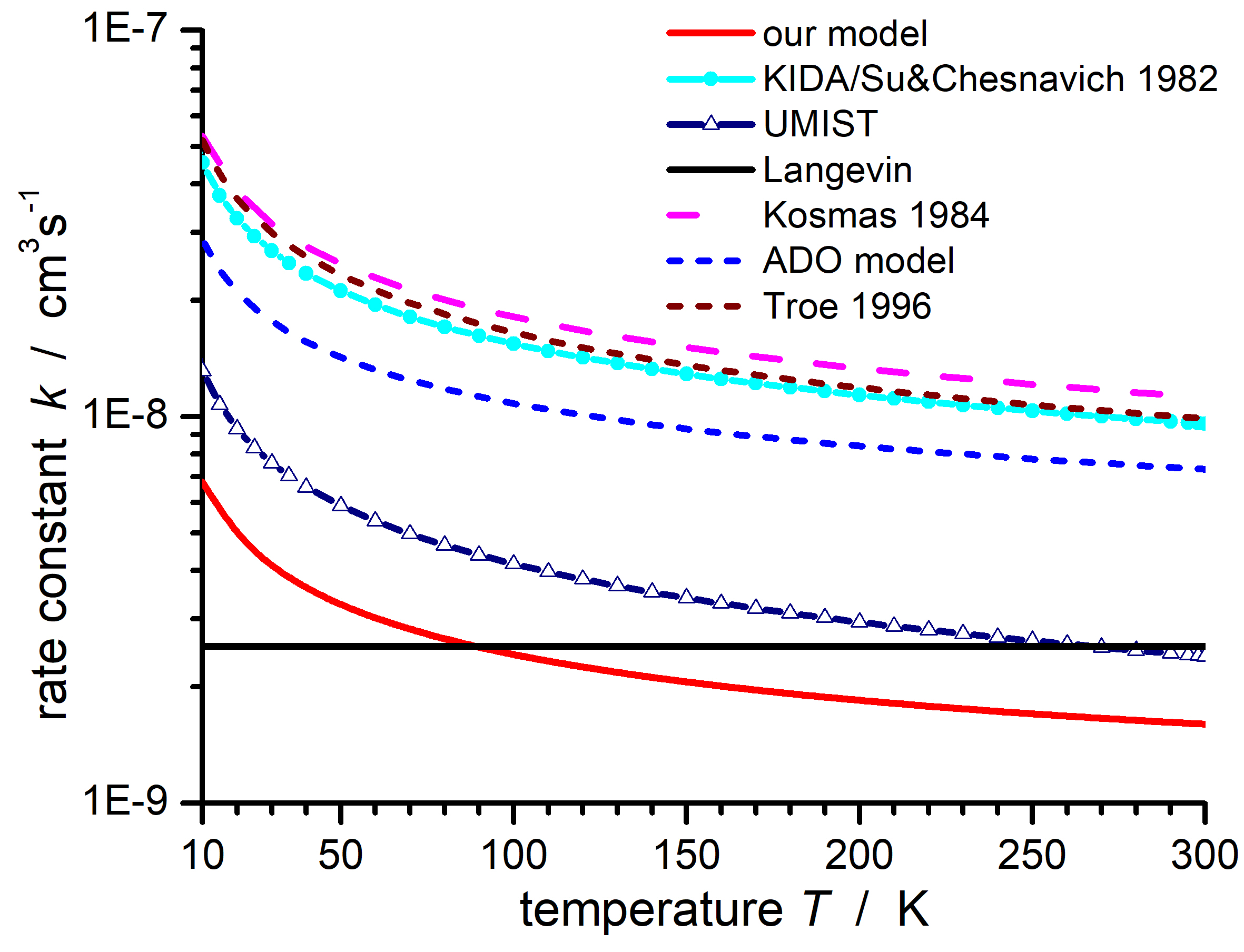}
  \caption{
  Total rate constants $k$ as a function of temperature $T$ for the reaction of 
  \ce{He^{+.}} with \ce{CH3CN}. \textit{Solid red line}: our model; \textit{solid black line}: Langevin model; \textit{cyan line with dots}: KIDA database \citep{KIDA2015, Su1982}; \textit{blue line with triangles} UMIST database \citep{UMIST2022}; \textit{dashed blue line}: scaled averaged dipole orientation (ADO) with $c=\overline{cos\theta}$=0.253 \citep{Su1975, Su1978}; \textit{dashed magenta line}: average $cos\theta$ method \citep{Kosmas1984}; \textit{dashed brown line}: statistical adiabatic channel model (SACM), eq. (5.2) from \citep{Troe1996}. 
  }
  \label{fig:rates}
 \end{figure}

From the CSs obtained using the approach described above it is possible to estimate the reaction rate coefficients as a function of temperature $k(T)$ by averaging the computed total CSs over a Maxwell-Boltzmann distribution of collision energies, as detailed in \citep{CeccarelliAscenzi2019}. Results are presented, with a solid red line, in Figure \ref{fig:rates} where they are compared with values from various astrochemical databases and a range of capture models.

The rates obtained using the pure Langevin model (\textit{i.e.} omitting the ion - permanent dipole interaction) is given by the solid black line, while the cyan line is from the KIDA database \citep{KIDA2015} that uses the expression from the Su and Chesnavich model derived from an empirical fit of variational transition state theory and classical trajectory calculations for point charged ions and polar neutrals \citep{Su1982}. The blue line with triangles is from the UMIST database \citep{UMIST2022} that proposes an expression for $k(T)$ assuming a value of $k$ at $T$=300 K equal to 
1.2x10$^{-9}$ cm$^{3}\cdot$s$^{-1}$, taken from \citep{Prasad}. The dashed blue line is from the scaled averaged dipole orientation (ADO) model and successive variants (already discussed in Sec. \ref{sec:expres1}) with a $c$ value of 0.253. It should be noted that this $c$ value refers to an average orientation calculated at 300 K in \citep{Su1975, Su1978}.
The dashed magenta line is from the average $cos\theta$ method \citep{Kosmas1984} and the dashed brown line corresponds to an analytical expression for the rate constant derived from the statistical adiabatic channel model (SACM) developed by Troe, more specifically from eq. (5.2) in \citep{Troe1996}.

Interestingly, in a recent experiment, the translational temperature dependence of the reaction rate constant for the charge exchange process between \ce{Ne^{+.}} ions and \ce{CH3CN} was measured at low temperatures \citep{Okada2020}, with the smaller measured reaction probability with respect to typical ion–polar molecule reactions at temperatures above 5 K being consistent with the $k(T)$ values obtained with our model for the case of \ce{He^{+.}} ions.

Previous astrochemical models, which used the reactions rates from the KIDA database, suggested that the \ce{CH3CN} observed in disks is mostly formed on grains, as gas-phase chemistry cannot account for the large observed abundances \citep{Oberg2015, Loomis2018}. 
These studies argued that, if N-bearing molecules are mostly formed on grains and therefore reflect the ice composition, the abundance ratios of \ce{CH3CN} and smaller N-bearing compounds (\textit{e.g.} \ce{HCN}, \ce{HC3N}) in disks can be compared with those in comets to test whether planets and small bodies inherit (at least partially) their composition from the disk where they are formed.

Our study, however, indicate that the rate coefficient for the destruction of \ce{CH3CN} through collisions with \ce{He^{+.}} evaluated at 10 K might be as low as $\sim6.8\times 10^{-9}$ cm$^{3}\cdot$molecule$^{-1}\cdot$s$^{-1}$, \textit{i.e.} almost one order of magnitude smaller than what currently reported in KIDA \citep{KIDA2015}. Combined with the revised chemical network for formation of \ce{CH3CN} \citep{Giani2023}, this suggests that, in contrast with what has previously been proposed, methyl cyanide in disks could be  mostly the product of gas-phase processes. The formation of abundant \ce{CH3CN} in the gas-phase would explain the routine detection of \ce{CH3CN} in planet-forming disks out to large radii (a few hundred of au), \textit{i.e.} out of the \ce{CH3CN} snowline where ices cannot be thermally evaporated (see \citet{Ceccarelli2023} and references therein for the binding energy of \ce{CH3CN}). 
This is in contrast with \ce{CH3OH}, which is only formed on grain surfaces at low temperatures via CO freeze-out and subsequent hydrogenation reactions \citep{watanabe2002}, and so is expected to be detected only where grains are sublimated, \textit{i.e.} inside the \ce{CH3OH} snowline. In fact, \ce{CH3OH} is only detected in a few disks, \textit{e.g.} in transition disks with an inner cavity \citep{Booth2021,booth2023}, or in outbursting disks \citep{lee_je2019}, where the snowline is pushed outwards. The low amount of methyl cyanide measured in comets compared to the larger values found for methanol (0.008 to 0.054\%  \textit{vs} 0.7 to 6.1 \% with respect to water \citep{Biver2019} may further support this hypothesis.
Our findings are therefore expected to have important implications for the origin of methyl cyanide in planet-forming disks, as well as for the comparison of the abundance ratios of N-bearing molecules in disks and in comets.

\section{Conclusions}
\label{sec:conclusions}
Guided ion beam experiments have been used to determine the collision energy dependence of total ($\sigma_T$) and partial CSs as well as BRs for the reaction of \ce{He^{+.}} with \ce{CH3CN}. While the collision energy, $E$, has been varied by about three orders of magnitude (from thermal up to hyper-thermal conditions), $\sigma_T$ (E) remains approximately constant (changing by only a factor two). 

These experimental findings contrast with predictions from widely-used capture models. In particular, the $\sigma_T$ (E) values predicted by capture models exhibit a sharper decrease with increasing $E$ than is observed experimentally. This is attributed to the fact that such models do not take into account the basic features of the crossing points between entrance and exit channels, which occur at much shorter distances than the maximum of the centrifugal barrier, and where non-adiabatic effects are triggered. Therefore, the possibility of reaching such crossings in the entrance channel, combined with the selective influence of their features and the centrifugal barrier, strongly limits the values and effectiveness of the orbital angular momentum, $l$, in promoting the reaction.
At high $E$, the probed intermolecular distances decrease and our treatment and the capture models tend to converge. 
This is because capture models do not consider the features of the crossings controlled by the interaction anisotropy, as they only evaluate the ion-permanent dipole interaction component. However, this omission is largely offset by the fact that they also do not account for the influence of the \ce{CH3} group's size.

To rationalize the experimental findings we have formulated detailed PESs for  the entrance and exit channels of the reaction. The PES in the entrance channel exhibits an high degree of anisotropy, as the interaction is strongly modulated by the molecular orientation with respect to \ce{He^{+.}}. During the collisions, the \ce{CH3CN} rotational motion perpendicular to its C$_{3v}$ symmetry axis, evaluated under thermal conditions typical of ion guided experiments, is hindered by comparatively high interaction potential energy barriers. In particular, at $R=20$ \AA, the ratio between the effective barrier height and the average rotational energy amounts to a factor 2, increasing up to a factor $\sim35$ at $R=5$ \AA\ and even higher for lower $R$. The polar \ce{CH3CN} molecule therefore adopts a “natural” orientation in the electric field gradient associated with the strong  anisotropic interaction potential. The orientation effect is expected to increase with decreasing $E$, becoming more effective than what observed in less anisotropic systems \citep{gisler2012probing, li2014communication, falcinelli2023role, cappelletti2024dawn, Willitsch_NatChem2024}. 

Therefore, the present investigation casts light on important stereo-dynamical effects whose role becomes prominent under thermal and sub-thermal conditions. In particular, decreasing $E$ the polar \ce{CH3CN} molecule tends to be canalized in restricted angular cones confined around $\theta=0\degree$, where the PES has the maximum attraction. However, under such conditions, the effective crossings between entrance and exit channels manifest at larger $R$ values, where the reduced overlap between orbitals involved in the electron transfer strongly limits the reactivity.
Our treatment provides rate coefficients expected to be reliable, within a factor 2-3, down to cold conditions (T $\sim$ 10 K) where our results represent a lower limit of the true values. The main uncertainty arises from neglecting, in the semiclassical calculation of CSs, quantum effects (\textit{e.g.} tunnelling trough the centrifugal barrier), whose role strongly increases with decreasing $T$. 

Our results for the reaction rates appear to be significantly different from those currently adopted in astrochemical network databases. On these basis, we propose that methyl cyanide in protoplanetary disks could be mostly the product of gas-phase processes. In conclusion it should be stressed that a complete characterisation of the chemical reactions, and theirreaction rates, leading to the formation and destruction of \ce{CH3CN} is crucial for the updating of astrochemical networks and the meaningful comparison of the chemical composition of disks and comets. 


\section*{Author Contributions}
\textbf{L. Mancini}: Software, Validation, Investigation, Visualization, Writing - Original Draft;  \textbf{E. V. Ferreira de Aragão}: Software, Validation, Investigation, Visualization; \textbf{F. Pirani}: Conceptualization, Methodology, Validation, Supervision, Formal analysis, Writing - Review \& Editing; \textbf{M. Rosi}: Software, Validation, Supervision, Resources; \textbf{N. Faginas-Lago}: Software, Validation, Supervision, Funding acquisition, Resources; \textbf{V. Richardson}: Investigation, Formal analysis, Writing - Review \& Editing; \textbf{L.M. Martini}: Supervision, Resources, Funding acquisition; \textbf{L. Podio}: Investigation, Conceptualization, Writing - Review \& Editing; \textbf{M. Lippi}: Investigation, Conceptualization,  Writing - Review \& Editing; \textbf{C. Codella}: Investigation, Conceptualization,  Writing - Review \& Editing; \textbf{D. Ascenzi}: Conceptualization, Methodology, Validation, Supervision, Formal analysis, Funding acquisition, Resources, Writing - Review \& Editing

\section*{Conflicts of interest}
There are no conflicts to declare.

\section*{Acknowledgements}
We thank Xiao He and Pengxiao Liang for their contribution to measurements. This work is supported by the European Union’s Horizon 2020 research and innovation programme under the Marie Sklodowska Curie grant agreement No 811312 for the project ”Astro-Chemical Origins” (ACO) and by  MUR PRIN 2020 project n. 2020AFB3FX  "Astrochemistry beyond the second period elements". D.A. also acknowledges financial support from the National Recovery and Resilience Plan (NRRP), Mission 4, Component 2, Investment 1.1, Call for tender No. 1409 published on 14.9.2022 by the Italian Ministry of University and Research (MUR), funded by the European Union – NextGenerationEU– Project Title P20223H8CK "Degradation of space-technology polymers by thermospheric oxygen atoms and ions: an exploration of the reaction mechanisms at an atomistic level" - CUP E53D23015560001. Views and opinions expressed are however those of the author(s) only and do not necessarily reflect those of the European Union or European Commission. Neither the European Union nor the granting authority can be held responsible for them. V.R. acknowledges funding for a PhD fellowship from the Dept. Physics, University of Trento. L.M. acknowledges funding from the European Union - NextGenerationEU under the Italian Ministry of University and Research (MUR) National Innovation Ecosystem grant ECS00000041 - VITALITY - CUP J97G22000170005


\bibliography{bibCH3CN} 
\bibliographystyle{aa} 

\end{document}